\def\aap{\ {A\&A}\ }

\def\apj{\ {ApJ}\ }

\def\apjs{\ {ApJS}\ }

\def\araa{\ {ARA\&A}\ }

\def\ban{\ {BAN}\ }

\def\mnras{\ {MNRAS}\ }

\def\pasp{\ {PASP}\ }
\def\pasj{\ {Publ. Astr. Soc. Japan}\ }

\documentclass[]{elsart}
\usepackage{natbib}
\usepackage{epsfig}

\newenvironment{mylisting}
{\begin{list}{}{\setlength{\leftmargin}{1em}}\item\scriptsize\bfseries}
{\end{list}}

\newcommand{\thost}{\mbox{$t_{\rm host}$}}
\newcommand{\tDP}{\mbox{$t_{\rm GPU}$}}
\newcommand{\fDP}{\mbox{$f_{\rm GPU}$}}
\newcommand{\tHA}{\mbox{$t_{\rm force}$}}
\newcommand{\tcomm}{\mbox{$t_{\rm comm}$}}
\newcommand{\tsend}{\mbox{$t_{\rm send}$}}
\newcommand{\nsend}{\mbox{$n_{\rm send}$}}
\newcommand{\tpred}{\mbox{$t_{\rm pred}$}}
\newcommand{\tcorr}{\mbox{$t_{\rm corr}$}}
\newcommand{\tbus}{\mbox{$t_{\rm bus}$}}
\newcommand{\nblock}{\mbox{$n_{\rm block}$}}
\newcommand{\npipe}{\mbox{$n_{\rm pipe}$}}
\newcommand{\tstep}{\mbox{$t_{\rm step}$}}
\newcommand{\nsteps}{\mbox{$n_{\rm steps}$}}
\newcommand{\trec}{\mbox{$t_{\rm rec}$}}
\newcommand{\nrec}{\mbox{$n_{\rm rec}$}}

\newcommand{\jerk}{\mbox{${\bf k}$}}

\def\apgt{\ {\raise-.5ex\hbox{$\buildrel>\over\sim$}}\ } 
\def\aplt{\ {\raise-.5ex\hbox{$\buildrel<\over\sim$}}\ } 
\def\lt{\ {\raise-.5ex\hbox{$\buildrel>$}}\ } 
\def\gt{\ {\raise-.5ex\hbox{$\buildrel<$}}\ }

\begin{document}

\begin{frontmatter}

%[HPC $N$-body simulations on GPU and GRAPE]
\title{High Performance Direct Gravitational N-body Simulations on
       Graphics Processing Units}

\author[SCS,API]{Simon F. Portegies Zwart}
\author[SCS]{Robert G. Belleman}
\author[SCS]{Peter M. Geldof}

\address[SCS]{Section Computational Science, University of Amsterdam, Amsterdam, The Netherlands}
\address[API]{Astronomical Institute "Anton Pannekoek" , University of Amsterdam, Amsterdam, The Netherlands}

\begin{abstract}

We present the results of gravitational direct $N$-body simulations
using the commercial graphics processing units (GPU) NVIDIA Quadro
FX1400 and GeForce 8800GTX, and compare the results with GRAPE-6Af
special purpose hardware. The force evaluation of the $N$-body problem
was implemented in Cg using the GPU directly to speed-up the
calculations.  The integration of the equations of motions were,
running on the host computer, implemented in C using the 4th order
predictor-corrector Hermite integrator with block time steps.  

We find that for a large number of particles ($N \apgt 10^4$) modern
graphics processing units offer an attractive low cost alternative to
GRAPE special purpose hardware.  A modern GPU continues to give a
relatively flat scaling with the number of particles, comparable to
that of the GRAPE.  Using the same time step criterion the total
energy of the $N$-body system was conserved better than to one in
$10^6$ on the GPU, which is only about an order of magnitude worse
than obtained with GRAPE. For $N\apgt 10^6$ the GeForce 8800GTX was
about 20 times faster than the host computer.  Though still about an
order of magnitude slower than GRAPE, modern GPU's outperform GRAPE in
their low cost, long mean time between failure and the much larger
onboard memory; the GRAPE-6Af holds at most 256k particles whereas the
GeForce 8800GTF can hold 9 million particles in memory.

\end{abstract}
\end{frontmatter}

\section{Introduction}

Since the first large scale simulations of self gravitating systems
the direct $N$-body method has gained a solid footing in the research
community. At the moment $N$-body techniques are used in astronomical
studies of planetary systems, debris discs, stellar clusters, galaxies
all the way to simulations of the entire universe
\citep{2006astro.ph..1232H}. Outside astronomy the main areas of
research which utilise the same techniques are molecular dynamics,
elementary particle scattering simulations, plate tectonics, traffic
simulations and chemical reaction network studies. In the latter
non-astronomical applications, the main force evaluating routine is
not as severe as in the gravitational $N$-body simulations, but the
backbone simulation environments are not very different.

The main difficulty in simulating self gravitating systems is the lack
of antigravity, which results in the requirement of global
communication; each object feels the gravitational attraction of any
other object.

The first astronomical simulation of a self gravitating $N$-body
system was carried out by \cite{1941ApJ....94..385H} with the use of
37 light bulbs and photoelectric cells to evaluate the forces on the
individual objects.  Holmberg spent weeks in order to perform this
quite moderate 37-particle simulation.  Over the last 60 or so years
many different techniques have been introduced to speed up the kernel
calculation. Today, such a calculation requires about 50\,000
integration steps for one dynamical time unit. At a speed of $\sim
10$\,Gflop/s the calculation would be performed in a few seconds.
% 1 GLOP corresponds to about 20000 steps/s

The gravitational $N$-body problem has made enormous advances in the
last decade due to algorithmic design.  The introduction of digital
computers in the arena
\citep{1963ZA.....57...47V,1964ApNr....9..313A,1968BAN....19..479V} led
to a relatively quick evaluation of mutual particle forces. Advanced
integration techniques introduced to turn the particle forces in a
predicted space-time trajectory, opened the way to predictable
theoretical results \citep{1975ARA&A..13....1A,1999PASP..111.1333A}.
One of the major developments in the speed-up and improved accuracy of
the direct $N$-body problem was the introduction of the block-time
step algorithm \citep{1991ApJ...369..200M,1993ApJ...414..200M}.

In the late 1980s it became quite clear that the advances of modern
computer technology via Moore's law \citep{Moore} was insufficient to
simulate large star clusters by the new decade
\citep{1988ApJS...68..833M,1990ApJ...365..208M}. This realization
brought forward the initiatives employed around the development of
special hardware for evaluating the forces between the particles
\citep{1986LNP...267...86A,1996IAUS..174..141T,1998sssp.book.....M,2001ASPC..228...87M,2003PASJ...55.1163M}, 
and of the efficient use of assembler code on general purpose hardware
\citep{2006NewA...12..169N,2006astro.ph..6105N}.

One method to improve performance is by parallelising force evaluation
Eq.\,\ref{Eq:Force} for use on a Beowulf or cluster computer (with or
without dedicated hardware)\citep{2006astro.ph..8125H}, a large
parallel supercomputer \citep{2002NewA....7..373M,2003JCoPh.185..484D}
or for grid operations \citep{2004astro.ph.12206G}.  In particular for
distributed hardware it is crucial to implement an algorithm that
limits communication as much as possible, otherwise the bottleneck
simply shifts from the force evaluation to interprocessor
communication.

A breakthrough in direct-summation $N$-body simulations came in the
late 1990s with the development of the GRAPE series of
special-purpose computers \citep{1998sssp.book.....M}, which achieve
spectacular speedups by implementing the entire force calculation in
hardware and placing many force pipelines on a single chip.  The
latest special purpose computer for gravitational $N$-body
simulations, GRAPE-6, performs at a peak speed of about 64\,Tflop/s
\citep{2001ASPC..228...87M}.

In our standard setup, one GRAPE-6Af processor board is attached to a
host workstation, in much the same way that a floating-point or
graphics accelerator card is used.  We use a smaller version: the
GRAPE-6Af which has four chips connected to a personal workstation via
the PCI bus delivering a theoretical peak performance of $\sim 131$
Gflops for systems of up to 128k particles at a cost of $\sim\$6$K
\citep{2005PASJ...57.1009F}.  Advancement of particle positions
[$\mathcal{O}(N)$] is carried out on the host computer, while
interparticle forces [$\mathcal{O}(N^2)$] are computed on the GRAPE.

The latest developments in this endeavour is the design and
construction of the GRAPE-DR, the special purpose computer which will
break the Pflop/s barrier by the summer of 2008
\citep{2005astro.ph..9278M}\footnote{See {\tt
http://grape.astron.s.u-tokyo.ac.jp/grape/computer/grape-dr.html}}.
One of the main arguments to develop such a high powered and
relatively diverse computer is to perform simulations of entire
galaxies \citep{2005JKAS...38..165M,Hoekstra}.

The main disadvantages of these special purpose computers, however, are
the relatively short mean time between failure, the limited
availability, the limited applicability, the limited on-board memory
to store particles, the simple fact that they are basically build by a
single research team led by prof. J. Makino and the lack of competing
architectures.

The gaming industry, though not deliberately supportive of scientific
research, has been developing high power parallel vector processors
for performing specific rendering applications, which are in
particular suitable for boosting the frame-rate of games.  Over the
last 7 years graphics processing units (GPUs) have evolved from fixed
function hardware for the support of primitive graphical operations to
programmable processors that outperform conventional CPUs, in
particular for vectorizable parallel operations. Regretfully, the
precision of these processors is still 32-bit IEEE which is below the
average general purpose processor, but for many applications it turns
out that the higher (double) precision is not crucial or can be
emulated at some cost.  It is because of these developments, that more
and more people use the GPU for wider purposes than just for graphics
\citep{GPUGems1,GPUGems2,buck04brook}. This type of programming is also
called general purpose computing on graphics processing units
(GPGPU)\footnote{see {\tt http://www.gpgpu.org}}.  Earlier attempts to
use a GPU for gravitational N-body simulations were carried out
approximate force evaluation methods using shared time steps
\cite{Nyland04}, but provide little improvement in performance.
A 25-fould speed increase compared to an Intel Pentium IV processor
was reported by \cite{Elsenetal06}, but details of their implemtation
of the force evaluation alrgorithm are yet unclear.

Using the GPU as a general purpose vector processor works as follows.
Colours in a computer are represented by one or more numbers. The
luminance can be represented by just a single number, whereas a
coloured pixel may contain separate values indicating the amount of
red, green and blue. A fourth value alpha may be included to indicate
the amount of transparency. Using this information, a pixel may be
drawn. There are many pixels in a frame, and ideally, these should be
updated all at the same time and at a rate exceeding the response time
of the human eye.  This requires fast computations for updating the
pixels, for example when a camera moves or a new object comes into
view. Such operations usually have an impact on many or even all
pixels fast computations are required.  But since the majority of
pixes do not require information from other pixes, processing can be
done efficiently in parallel. All information required to build a
pixel should go through a series of similar operations, a technique
which is better known as single instruction, multiple data
(SIMD). There are many different kinds of operations this information
needs to go through. The stream programming model has been designed to
make the information go through these operations efficiently, while
exposing as much parallelism as possible. The stream programming model
views all informations as ``streams'' of ordered data of the same data
type. The streams pass through ``kernels'' that operate on the streams
and produce one or more streams as output.

In this paper we report on our endeavour to convert a high precision
production quality $N$-body code to operate with graphics processor
units. In \S\,\ref{Sect:Nbody} we explain the adopted $N$-body
integration algorithm, in \S\,\ref{Sect:Cg} we address the programming
environment we used to program the GPU, In the sections
\S\,\ref{Sect:Results} and \S\,\ref{Sect:performance} we present the
results on two GPUs and compare them with GRAPE-6Af and we discuss a
model to explain the GPUs performance. In
\S\,\ref{Sect:Discussion} we summarise our findings, and in the Appendix we present a snippet of the source code in Cg.

\section{Calculating the force and integrating the particles}
\label{Sect:Nbody}

The gravitational evolution of a system consisting of $N$ stars with
masses $m_j$ and at position ${\bf r}_j$ is computed by the direct
summation of the Newtonian force between each of the $N$ stars.  The
force ${\bf F}_i$ acting on particle $i$ is then obtained by summation
of all other $N-1$ particles
\begin{equation}
  {\bf F}_i \equiv m_i {\bf a}_i =     m_i G \sum^{N}
                   _{j=1, j \ne i} 
                   m_j 
                  {{\bf r}_i-{\bf r}_j \over |{\bf r}_i-{\bf r}_j|^3}.
\label{Eq:Force}\end{equation}
Here $G$ is the Newton constant.

A cluster consisting of $N$ stars evolves dynamically due to the
mutual gravity of the individual stars. For an accurate force
calculation on each star a total of ${1 \over 2} N(N-1)$ partial
forces have to be computed.  This {\large{O}($N^2$)} operation is the
bottleneck for the gravitational $N$-body problem.  

The GPU scheme described in this paper is implemented in the 
$N$-body integrator. Here particle motion is calculated using a
fourth-order, individual-time step ``Hermite'' predictor-corrector
scheme (Makino and Aarseth 1992).\nocite{1992PASJ...44..141M} This
scheme works as follows.  During a time step the positions (${\bf x}$)
and velocities (${\bf v} \equiv \dot{{\bf x}}$) are first predicted to
fourth order using the acceleration (${\bf a} \equiv \ddot{{\bf x}}$)
and the ``jerk'' ($\jerk \equiv \dot{{\bf a}}$, the time derivative
of the acceleration) which are known from the previous step.

The predicted position (${\bf x}_p$) and velocity (${\bf v}_p$) are
\begin{eqnarray}
        {\bf x}_p &=& {\bf x} + ({\bf v} + (dt/2)  ({\bf a} + (dt/3)  \jerk)) dt, \\
        {\bf v}_p &=& {\bf v} + ({\bf a} + (dt/2)  \jerk) dt.
\end{eqnarray}

The acceleration and jerk are then recalculated at the predicted time,
using $x_p$ and $v_p$. Finally, a correction is based on the estimated
higher-order derivatives:
\begin{eqnarray}
        {\bf a3} &=&  2  ({\bf a} - {\bf a}_p) + (\jerk + \jerk_p) dt, \\
        {\bf a2} &=& -3  ({\bf a} - {\bf a}_p) - (2 \jerk + \jerk_p) dt.
\end{eqnarray}
where 
\begin{eqnarray}
        {\bf a2}   &=&  \dot{\jerk}  dt^2 / 2, \\       
        {\bf a3}   &=&  \ddot{\jerk}  dt^3 / 6.
\end{eqnarray}
Which then leads to the new position and velocity at time $t+dt$. 
\begin{eqnarray}
        {\bf x} &=& {\bf x}_p + ({\bf a2}/12 + {\bf a3}/20)  dt^2, \\
        {\bf v} &=& {\bf v}_p + ({\bf a2}/3 + {\bf a3}/4)  dt.
\end{eqnarray}

The new ${\bf a}$ and $\jerk$ are computed by direct summation, and
the motion is subsequently corrected using the additional derivative
information thereby obtained.

A single integration step in the integrator proceeds as follows:
\begin{itemize}
\item[$\bullet$] Determine which stars are to be updated. Each star
      has an individual time ($t_i$) associated with it at which it
      was last advanced, and an individual time step ($dt_i$). The
      list of stars to be integrated consists of those with the
      smallest $t_i+dt_i$.  Time steps are constrained to be powers of
      2, allowing ``blocks'' of many stars to be advanced
      simultaneously \citep{1993ApJ...414..200M}.
\item[$\bullet$] Before the step is taken, check for system
     reinitialization, diagnostic output, termination
     of the run, storing data.
\item[$\bullet$] Perform low-order prediction of all particles to the
      new time $t_i+dt_i$. This operation may be performed on the
      GPU, if available.
\item[$\bullet$] Recompute the acceleration and jerk on all stars in
      the current block (using the GPU, if available), and correct
      their positions and velocities to fourth-order.
\end{itemize}

Note that this scheme is rather simple as it does not include
treatment for close encounters, binaries or higher order (hierarchical
or democratic) stable multiple systems.

\begin{table}
\caption[]{ Detailed information on the hardware used in our
experiments.  The first column gives the parameter followed by the
four different hardware setups (GRAPE-6Af, GeForce 8800GTX, Quadro
FX1400 and information about the host computer.  The information for
the GRAPE is taken from \cite{2003PASJ...55.1163M}, the GPU
information is from {\tt http://www.nvidia.com}.
The hardware details are the number of processors pipelines (\npipe),
the processor's clock frequency ($\fDP \equiv 1/\tDP$), the memory
bandwidth for communication between host and attached processor
($1/\tbus$), the amount of memory (in number of particles, one particle
requires 84\,bytes, here we adopt $1{\rm k} \equiv 1024$).  For measured
hardware parameters, see Tab.\,\ref{Tab:Hardware}.
\label{Tab:GPU}
}
\begin{tabular}{lrrrrr}
data                     & GRAPE-6Af & 8800GTX & FX 1400& Xeon & unit \\
\hline				
\npipe                   &    48     & 128     & 12     &   1   & \\
\fDP                     &    90     & 575     & 350    &  3400 & MHz  \\
$1/\tbus$                &    33.8   & 86.4    & 19.2   &  NA   & GB/s\\
%Memory                  &    11.0   & 768     & 128    &  --   & Mbyte \\
Memory                   &    128k   & 9362k   & 1562k  &  --   & particles \\
%\hline
%$\eta_{\rm clock}$       &    1      & 75      & 25     &  1    & cycles \\
%$t_{\rm overhead}$       &    0.04   & 1.0     & 1.0    &  NA   & ms \\
%%Shader Clock             &         & 1350    &        & MHz \\
%%Memory Clock             &         & 900     &        & MHz \\
%%Memory Amount            &         & 768     & 128    & Mb \\
%%Memory Interface         &         & 384     & 256    & bit \\
\hline
\end{tabular}
\end{table}

\section{The programming environment}\label{Sect:Cg}

The part of the algorithm that executes on the GPU (the force
evaluation) is implemented in the Cg computer language (C for
graphics, \cite{Cg}, see Appendix A), which has a syntax quite similar
to C. The Cg programming environment includes a compiler and run-time
libraries for use with the open graphics library (OpenGL)\footnote{see
{\tt http://www.opengl.org}} and DirectX\footnote{see {\tt
http://www.microsoft.com/directx}} graphics application programming
interfaces.  Though originally developed for the creation of real-time
special effects without the need to program directly to the graphics
hardware assembly language, researchers soon recognised the potential
of Cg and started to apply it not only to high-performance graphics
but also to a wide variety of ``general-purpose computing'' problems
\citep{GPUGems1,GPUGems2}.

\subsection{Mapping the $N$-body problem to a GPU}\label{Sect:Mapping}

The challenge in the implementation of an efficient $N$-body code on a
GPU lies in the mapping of the algorithm and the data to 
graphical entities supported by the Cg language.  Particle data arrays
are represented as ``textures''. Normally, textures are used to
represent pixels colour attributes with one single component
(luminance, red, green, blue or alpha), three components (red, green
and blue) or four components (red, green, blue and alpha). In our
implementation we use multiple textures to represent the input and
output data of $N$ particles, as follows:
\begin{itemize}
  \item{} Input: mass ($N$), position ($3N$) and velocity ($3N$)
  \item{} Output: acceleration ($3N$), jerk ($3N$) and potential ($N$)
\end{itemize}
All values are represented as single precision (32-bit) floating point
numbers for a total of 21 floats or 84\,bytes per particle. In Appendix
A we present a snippet of the source code in Cg, showing the
implemented force evaluation routine.  With the 768\,Mbyte on-board
memory of the GeForce 8800GTX it can store about 9 million particles,
whereas the GRAPE-6Af can store only 128k (see Tab.\,\ref{Tab:GPU}).

Transferring data from CPU to GPU is accomplished through the
definition of textures, which can either be read-only or write-only,
but not both at the same time. The data structures in the CPU are then
copied onto appropriately defined textures in the graphics card's
memory.  Obtaining the results from GPU to CPU is done by reading back
the pixels from the appropriate rendering targets into data structures
on the host CPU.  Therefore the output textures (acceleration, jerk
and potential) are represented by a double-buffered scheme, where
after each GPU computation the textures are swapped between reading
and writing. There is some additional overhead (of order $N$) for this operation which has to be performed every block time
step.

Conventionally, graphics cards render into a ``frame buffer'', a
special memory area that represents the image seen on a
display. However, a frame buffer is unsuitable for our purposes as the
data elements in this buffer are ``clamped'' to value ranges that map
the capabilities of the display.  Invariably this means that 32-bit
real vectors are reduced in resolution and therefore in accuracy
too. This is perfectly fine for visual displays where the number of
colours after clamping are still $2^{24}$ ($\approx 16$ million),
sufficient to make two neighbouring colours indiscernible to the human
eye. However, this is unacceptable for scientific production
calculations. The workaround is to create an off-screen frame buffer
object and instruct GPU programs to render into these rather than to
the screen. Off-screen frame buffers support 32-bit floating point
values and are not clamped and therefore preserve their precision.

The GPU has two main kernel operations available in programmable
graphics pipelines, these are a ``vertex shader'' and a ``fragment
shader''. Our implementation only makes use of the fragment shader
pipeline as it is better suited for the kind of calculations in the
N-body problem and because the fragment pipeline in general provides
more processing power\footnote{Before the 8800GTX family of GPUs,
vertex programs and fragment programs had to execute on distinct
processing units. The 8800GTX is the first generation of GPUs where
this distinction no longer exists and the two are unified.}.  The host
CPU is responsible for allocating the input textures and frame buffer
objects, copy the data between CPU and GPU, and binding textures that
are to be processed by kernels.  The lower order prediction and
correction of the particle positions is done on the host CPU. In
Tab.\,\ref{Tab:GPU} we summarise the hardware properties of the two
adopted GPU's and the GRAPE.

\section{Results}\label{Sect:Results}

\begin{table}
\caption[]{Results of the performance measurements for a Plummer
           sphere with $N$ equal mass particles initially in virial
           equilibrium for 0.25$N$-body time units (from $t = 0.25$ to
           $t=0.5$) using a softening of 1/256.  In the first column
           we list the number of particles, followed by the timing
           results of the GRAPE in seconds. In the last column we give
           the timing results for the calculation without an attached
           processor.  The GRAPE (second column) was measured up to
           128k particles, because the on-board memory did not allow
           for larger simulations. Simulations on the FX1400 and the
           host computer were limited for practical reasons.}
\label{Tab:Results}
\begin{tabular}{lccccc}
\hline
$N$       & GRAPE-6Af   & 8800GTX & FX1400   & Xeon \\
\hline
256       &   0.07098   &     2.708&   3.423  &     0.1325 \\
512       &   0.1410    &     8.777&   10.59  &     0.5941 \\
1024      &   0.3327    &    17.46 &   20.20  &     2.584  \\
2048      &   0.7652	&    45.27 &   54.16  &    10.59  \\
4096      &   1.991	&   128.3  &   157.8  &    50.40  \\
8192      &   5.552	&   342.7  &   617.3  &   224.7  \\
16384     &  16.32	&   924.4  &   3398   &   994.0  \\
32768     &  51.68	&  1907    &  13180   &  4328  \\
65536     & 178.2       &  3973    &  40560   & 19290  \\
131072    &  -          &  8844    & -        & - \\
262144    &  -          & 22330    & -        & - \\
524288    &  -          & 63960    & -        & - \\
\hline
\end{tabular}
\end{table}

To test the various implementations of the force evaluator we perform
several tests on different hardware.  For clarity we perform each test
with the same realization of the initial conditions.  For this we vary
the number of stars from $N=256$ particles with steps of two to half a
million stars (see Tab.\,\ref{Tab:Results}). Not each set of initial
condition is run on every processor, as the Intel Xeons, for example,
would take a long time and the scaling with $N$ is unlikely to change
as we clearly have reached the CPU-limited calculation regime (see
\S\,\ref{Sect:performance}).

The initial conditions for each set of simulations were generated by
randomly selecting the stellar positions and velocities according to
the \cite{1911MNRAS..71..460P} distribution using the method described
by \cite{1974A&A....37..183A}.  Each of the stars were given the same
mass.  The initial particle representations were scaled to virial
equilibrium before starting the calculation.

Each set of initial conditions is run from $t=0$ to $t=0.50$ time
units \citep{1986LNP...267..233H}\footnote{see also {\tt
http://en.wikipedia.org/wiki/Natural\_units\#N-body\_units}}, but the
performance is measured only over the last quarter of an $N$-body time
unit to reduce the overhead for reading the snapshot and the
initialisation of the integrator.  The maximum time step for the
particles was 0.125, to guarantee that each particle was evaluated at
least twice during the course of the simulation.  The force
calculations were performed by adopting a softening of $1/256$ for all
simulations.

For our performance measurements we have used four nodes of a
Hewlett-Packard xw8200 workstation with a dual Intel Xeon CPU running
at 3.4 GHz and either the GRAPE, a Quadro FX1400 or GeForce 8800GTX
graphics card in the PCI Express ($16\times$) bus.  The cluster nodes
were running a Linux SMP kernel version 2.6.16, Cg version 1.4,
graphics card driver version 1.0-9746 and the OpenGL 2.0 bindings.

\begin{figure}
\psfig{figure=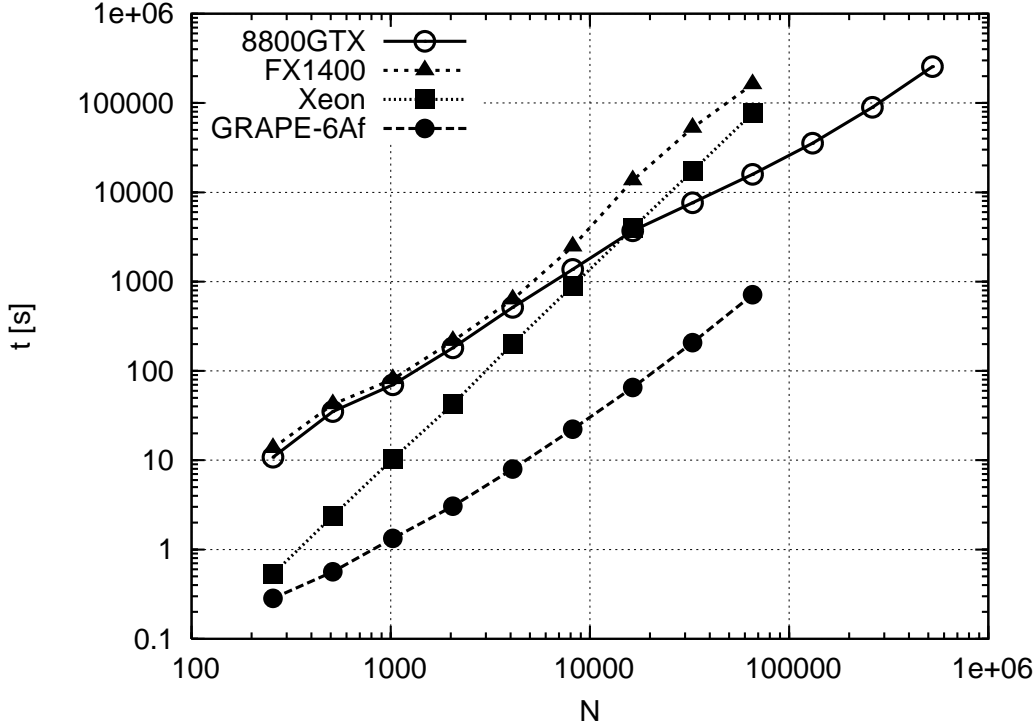,width=\columnwidth}
   \caption[]{ Timing of several implementations of the gravitational
$N$-body simulations for $N=256$ particles to $N=512$k particles (only
for the 8800GTX, the others up to 64k) over one $N$-body time unit.
The 8800GTX are represented with open circles connected with a solid
curve, the GRAPE is given by bullets with dashed line.  The thin
dashed (triangles) line and thin dotted (squares) lines give the
results of the calculations with the FX1400 and with only the host
computer. Note that in timings in Tab.\,\ref{Tab:Results} were
multiplied by a factor of four to estimate the compute time for one
dynamical time unit, rather than the $1/4^{\rm th}$ over which the
timing calculations were performed.
%
%GRAPE: dash + bullets
%GTX: solid + circles
%FX: dash + triangles
%PC: sqyuares + dotts
\label{fig:GPU} }
\end{figure}

In Figure\,\ref{fig:GPU} we show the timing results of the $N$-body
simulations.  The FX1400 is slower than the general purpose computer
over the entire range of $N$ in our experiments. The bad performance
of the FX1400 is mainly attributed to the additional overhead in
communication and memory allocation.  For $N \aplt 10^4$ the GeForce
8800GTX GPU is slower than the host computer but continues to
have a relatively flat scaling, comparable to the GRAPE-6, whereas the
host has a much worse ($\propto N^2$) scaling.  The scaling of the
compute time of the GPU is proportional to that of the GRAPE ($\propto
N^{3/2}$), but the latter has a smaller offset by about an order of
magnitude. This is mainly caused by the efficient use of the GRAPE
pipeline, which requires fewer clock cycles per force evaluation
compared to the GPU (see
\S\,\ref{Sect:Discussion}).

\section{Performance modelling of the GPU}\label{Sect:performance}

In modelling the performance of the GPU we adopt the model proposed by
\cite{2002NewA....7..373M,2006astro.ph..8125H} but tailored to the host plus 
GPU and to the GRAPE architecture.

The wall clock time required for advancing the \nblock\ particles in a
single block time step in the $N$-body systems is
\begin{equation}
  \tstep = \thost + \tHA + \tcomm. 
\end{equation}
Here $\thost = \tpred + \tcorr$ is the time spend on the host computer
for predicting and correcting the particles in the block, \tHA\, is
the time spend on the attached processor and \tcomm\, is the time
spend communicating between the host and the attached processor.  We
now discuss the characteristics of each of the elements in the
calculation for \tstep.

\paragraph{Host operation.}

The predictions and corrections of the particles are calculated on the
host computer, and the time for this operation is directly related to
the speed of the host processor $t_{\rm cpu}$, the number of
operations in the prediction step $n_{\rm pred}$ and in the correction
step $n_{\rm corr}$. The total time spend per block step then yields
\begin{equation}
  \tpred \simeq n_{\rm pred} t_{\rm cpu} N,
\end{equation}
for the prediction and 
\begin{equation}
  \tcorr \simeq n_{\rm corr} t_{\rm cpu} N.
\end{equation}
for the correction. The number of operations per prediction step
$n_{\rm pred} \simeq 300$ and for the correction $n_{\rm corr} \simeq
1000$. This operation could be performed on the GPU, though the GRAPE
is not designed for the predictior and corrector calculation. For a
fair comparison between the GRAPE and the GPU and in order to preserve
high accuracy we performed these calculations on the host.

\paragraph{Communication.}\label{Sect:Communication}
The time spend communicating between the host and the attached
processor is expressed by the sum of the time needed to send \nsend\,
particles to the acceleration hardware and the time needed to receive
\nrec\, particles from the acceleration hardware:
\begin{equation}
  \tcomm =  \eta_{\rm send} \tsend \nsend + \eta_{\rm rec} \trec \nsend.
\end{equation}
Here $\eta_{\rm send} \tsend$\, and $\eta_{\rm rec}\trec$\, are the
time needed to send and receive one particle, respectively.  For the
computation without the hardware acceleration $\tcomm = 0$, since the
forces between all particles are calculated locally.  For the GRAPE
and the GPUs, however, a considerable amount of time is spend in
communication. For the GRAPE sending data is equally fast as receiving
data, i.e: $\tsend = \trec = \tbus$.  Sending data to the GPU is
considerably slower than receive data (see Tab.\,\ref{Tab:Hardware}).

The two send and receive efficiency factors $\eta_{\rm send}$ and
$\eta_{\rm rec}$, are the product of the overhead $\eta_o$ and the
number of bytes per particle that has to be send or received.  The
overhead $\eta_o = 188$ \citep{2005PASJ...57.1009F} for each of the
attached processors. Since for the GRAPE the send and receive
operation are equally expensive we can just count the number of bytes
that has to be transported per particle, which for the GRAPE hardware
is 72\,bytes \citep{2005PASJ...57.1009F,2006astro.ph..8125H}.  For the
GRAPE we then write $\eta_{\rm send}\tsend + \eta_{\rm rec}\trec = 72
\times 188 \tbus$.

For the GPU $\eta_{\rm send} > \eta_{\rm rec}$ (see
Table\,\ref{Tab:Hardware} for the measured values). The additional
overhead $\eta_o$ is the same as for the GRAPE, but per particle the
number of bytes to send is different from than the number to receive.
As we discussed in \S\,\ref{Sect:Mapping} a total of 56\,bytes has to be
send from the host to the GPU, whereas only 28\,bytes are received.  For
the GPU we then write $\eta_{\rm send} = 56 \times 188$, whereas
$\eta_{\rm rec} = 28 \times 188$.

In addition to the difference in the speeds for sending and receiving
data, the GPU suffers from an additional penalty.  The GRAPE sends the
particles in the block ($\nsend = \nblock$), whereas due to internal
memory management the GPU has to send and receive all particles
$\nsend = N$ (see \S\,\ref{Sect:Mapping}). This efficiency loss is
quite substantial, and will probably be reduced when we use CUDA as
programming environment (see \S\,\ref{Sect:Discussion}).

For the adopted (Hermite predictor-corrector block time-step)
integration scheme the number of particles in a single block \nblock\,
cannot be determined implicitly, though theoretical arguments suggest
$\nblock \propto N^{2/3}$. Instead of using this estimate We fitted
the average number of particles in a block time step. This fit was
done with the equal mass Plummer sphere initial conditions running on
GRAPE and run over one dynamical (N-body) time unit. The average
number of particles in a single block is then
\begin{equation}
   \nblock \simeq 0.20 n^{0.81}.
\label{Eq:Nblock}
\end{equation}

\paragraph{Calculation.}
The time spend by the hardware acceleration (\tHA) is directly related
to the speed of the dedicated processor (\tDP), the number of
pipelines per processor (\npipe) and the number of operations for one
force evaluation ($\eta_{\rm fe} \simeq 60$).
\begin{equation}
  \tHA = \eta_{\rm fe} N \nblock \tDP / \npipe.
\end{equation}
The details of the different hardware are presented in
Tab.\,\ref{Tab:GPU} and the measured values are in
Tab.\,\ref{Tab:Hardware} The GRAPE has a vector pipeline for each
processor which allows a more efficient force evaluation than the
GPU's, the number of operations per force evaluation for the GRAPE
therefore $\eta_{\rm fe} \simeq$ {\rm O}(1).

In order to enable hardware acceleration on our $N$-body code we had
to introduce a number of additional operations, like reallocating
arrays, which give rise to an extra computation overhead. For the
calculations with the host computer without hardware acceleration we
adopt $\eta_{\rm fe} \simeq 180$, a factor of three larger than for the
GPUs.

\paragraph{Total performance.}
The total wall-clock time spent per dynamical (N-body) time unit is
then
\begin{equation}
  t = \nsteps \tstep.
\end{equation}
Here we fitted to number of block steps per dynamical (N-body) time
units.  According to \cite{1988ApJS...68..833M,1990ApJ...365..208M}
$\nsteps
\propto n^{1/3}$. We measured the number of block time steps using the
equal mass Plummer distributions as initial conditions, using the
GRAPE enabled code and fitted the result:
\begin{equation}
  \nsteps \simeq 247 N^{0.35}.
\end{equation}

\begin{figure}
\psfig{figure=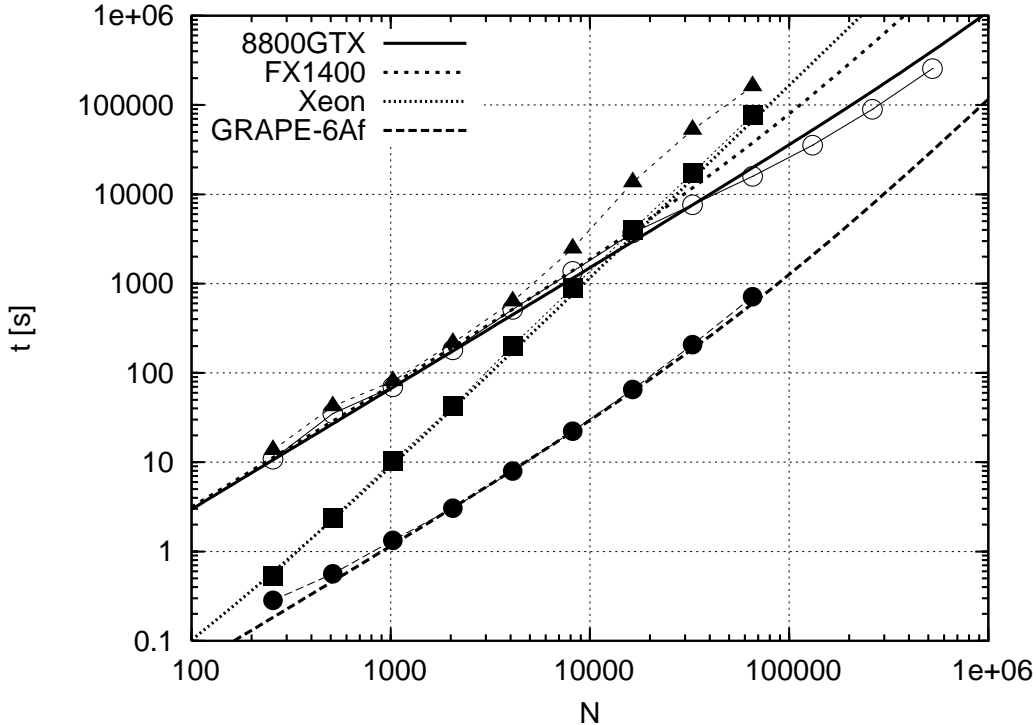,width=\columnwidth}
   \caption[]{The results of the above described performance model
   (thick lines) over-plotted with the results of the measurements for
   the three attached processors (symbols). The bullets represents the
   results from a single GRAPE-6Af processor board, the squares give
   the host workstation, the circles are for the GeForce 8800GTX and
   the triangles give the FX 1400 graphics processor.
\label{fig:PerformanceModel} }
\end{figure}

In fig.\,\ref{fig:PerformanceModel} we compare the results of the
performance model with the measurements on the workstation without
additional hardware (squares) and with three attached processors; a
single GRAPE-6Af processor board (bullets), an FX 1400 (triangles) and
the newer GeForce 8800GTX (circles). Note that the measurements in
Tab.\,\ref{Tab:Results} were multiplied by a factor four to compensate
for the fact that we performed our timings only over a quarter
$N$-body time unit.  Though these curves are not fitted, they give a
satisfactory comparison.

The largest discrepancy between the performance model and the
measurements can be noticed for the FX1400 GPU, which, for $N\apgt
10^4$ seems to perform considerably less efficiently than expected
according to the performance model. Part of this discrepancy, though
not explicitly mentioned in \S\,\ref{Sect:Communication} is in part
the result of a hysteresis effect in the communication of both
GPU's. For the 8800GTX, however, this effect is less evident, but
still present. Both GPU's tend to have a maximum communication speed
for blocks of 0.5\,Mbyte (about 6000 particles). An additional effect
which causing performance loss on the FX\,1400 is the increase in the
number of block time steps. This number continued to increase beyong
our measurements performed with GRAPE (see Eq.\,\ref{Eq:Nblock}).

The numbers listed in Tab.\,\ref{Tab:Hardware}, and used in our
performance model, are the optimum values.  The communication speed
drops by about a factor of two for much larger amounts of data
transfer to and from the GPU. For the FX1400, this drop in
communication is considerable, whereas for the 8800GTX it results in a
smaller performance loss (mainly due to the larger number of processor
pipelines).  The discrepancy for the GRAPE calculation with low $N$ is
the result of neglecting the limited size of the processor pipeline in
the performance model and due to the irregular behavior of the number
of particles in each block time step.

\begin{table}
\caption[]{
Measurements for the various hardware using in this paper.  The first
line gives the time spent by the host computer for predicting and
correcting a single particle. The second row is for calculating the
force between two particles. The last two columns give the time to
send a single particle to, and to receive a single particle from the
attached hardware. For the calculations with only the host computer
this operation is not available.  In particular the communication with
the GPUs turns out to be relatively slow.
\label{Tab:Hardware}
}
\begin{tabular}{lcccccc}
param & GRAPE-6Af            & 8800GTX & FX 1400  & Xeon  \\
\hline
\thost&$3.82 \times 10^{-7}$&$3.82 \times 10^{-7}$&$3.82 \times 10^{-7}$&$3.82 \times 10^{-7}$ \\
%$\eta_{\rm fe}\tHA$&$2.31 \times 10^{-10}$&$8.15 \times 10^{-10}$&$1.43 \times 10^{-8}$&$5.29 \times 10^{-8}$ \\
$\eta_{\rm fe}\tHA$&$1.11 \times 10^{-8}$&$1.04 \times 10^{-7}$&$1.72 \times 10^{-7}$&$5.29 \times 10^{-8}$ \\
$\eta_{\rm send}\tsend$&$2.00 \times 10^{-7}$ &$1.76 \times 10^{-5}$ &$1.89 \times 10^{-5}$&NA\\
$\eta_{\rm rec}\trec$ &$2.00 \times 10^{-7}$ &$5.97 \times 10^{-6}$ &$5.98 \times 10^{-6}$&NA\\
\hline
\end{tabular}
\end{table}

\begin{figure}
\psfig{figure=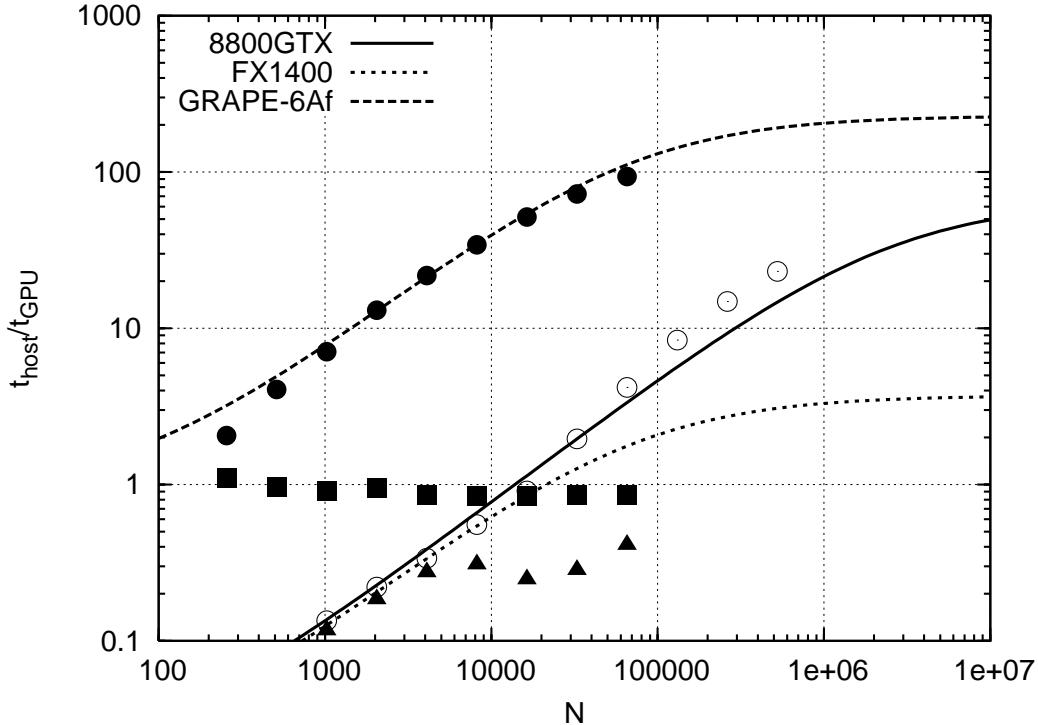,width=\columnwidth}
   \caption[]{The speedup of the two GPUs and the GRAPE with respect
   to the host workstation as a function of the number of particles.
   The (lower) dotted curve is for the Quadro FX1400, the solid curve
   (middle) gives the timing for the GeForce 8800GTX and the top line
   (dashes) represents the GRAPE.
\label{fig:Speedup} }
\end{figure}

In fig.\,\ref{fig:Speedup} we present the speed-up for the various
hardware configurations, compared to running on the host workstation.
Here it is quite clear that for low $N$ the GPU's do not give a
appreciable speedup, but for a large number of particles, the GeForce
8800GTX gives a speedup of at least an order of magnitude, but not as
much as the GRAPE. The latter, however, will not be able to perform
simulation of more than 128k particles\footnote{Due to a defective
chip on our GRAPE-6 the on-board memory was reduced from 128k particles
to 64k particles.  The latest GRAPE-6Af are equipped with 256k
particles of memory.}.

\section{Discussion}\label{Sect:Discussion}

We have successfully implemented the direct gravitational force
evaluation calculation using Cg on two graphics cards, the NVIDIA
Quadro FX1400 and the NVIDIA GeForce 8800GTX, and compared their
performance with the host workstation and the GRAPE-6Af special
purpose computer.

For $N \aplt 10^4$ objects the workstation outperforms the GPUs. This
is mainly due to additional overhead introduced by the communication
to the GPU and memory allocation on the GPU. For a larger number of
particles the more modern GPU (8800GTX) outperforms the workstation by
up to about a factor of 50 (for 9 million particles). Such a large
number of particles cannot be simulated on the GRAPE-6Af, due to
memory limitations. For up to 256k, the maximum number of particles
that can be stored on the GRAPE, the 8800GTX is slower than the GRAPE
by about an order of magnitude. Still, at this particle number the GPU
is faster than the workstation by an order of magnitude.

The GPU has lower accuracy compared to the GRAPE or the host
workstation.  The GRAPE-6 uses 24-bit mantissa for calculating the
differential position, and 64bit fixed point format for accumulation.
The pipeline for the time derivative is designed with 20bit mantissa
and 32 bit fixed-point notation for the final accumulation
\citep{2003PASJ...55.1163M}. In principle the NVIDIA architecture
should not be able to achieve similar precision, but would fall short
in precision by about an order of magnitude compared to the GRAPE-6.
The average mean error in the energy is $(1.7 \pm 1.6)
\times 10^{-6}$ for the 8800GTX and $(5.1 \pm 0.56) \times 10^{-6}$
for the FX1400 (averaged over the simulations for $N=256$ to $N=64$k),
whereas for the GRAPE we measured $(1.9 \pm 1.2) \times 10^{-7}$,
which is comparable to the mean error on the host. Both the host and
GRAPE produce an energy error which is about an order of magnitude
smaller than that of the GPUs.

The precision of the GPU is regretfully unlikely to increase anytime
soon, as the higher precision is not required for graphics
applications and it would imply a considerable redesign of the
hardware.  But we could improve the accuracy of the GPU even further
by sorting the forces on size before adding them, summing the smallest
forces first.

The fixed point notation in the GRAPE-6 allows for a much more
efficient use of clock cycles, allowing effectively one operation per
clock cycle, whereas the NVIDIA architecture requires more
cycles. This turns out to be an important reason why the 8800GTX is
slower than the GRAPE-6.

The main advantage of the GPU over that of the dedicated GRAPE
hardware, is the much larger memory, the wider applicability and the
much lower cost of the former. The large memory on the GPU allows
simulations of up to about 9 million particles, though one has to wait
for about two years for one dynamical time scale.

In theory the 8800GTX should be able to outperform the GRAPE-6Af, but
due to relatively inefficient memory access and additional overhead
cost, which is not present in the GRAPE hardware, many clock cycles
seem to be lost. With a more efficient use of the hardware the GPU
could, in principle, improve performance by about two orders of
magnitude. For the next generation of GPUs we hope that this
efficiency bottleneck will be lifted. In that case, the GPU would
outperform GRAPE by almost an order of magnitude.  Note, however, that
the GRAPE-6 is based on 5 year old technology, and the next generation
GRAPE is likely to outperform modern GPUs by a sizable margin.

\begin{figure}
\psfig{figure=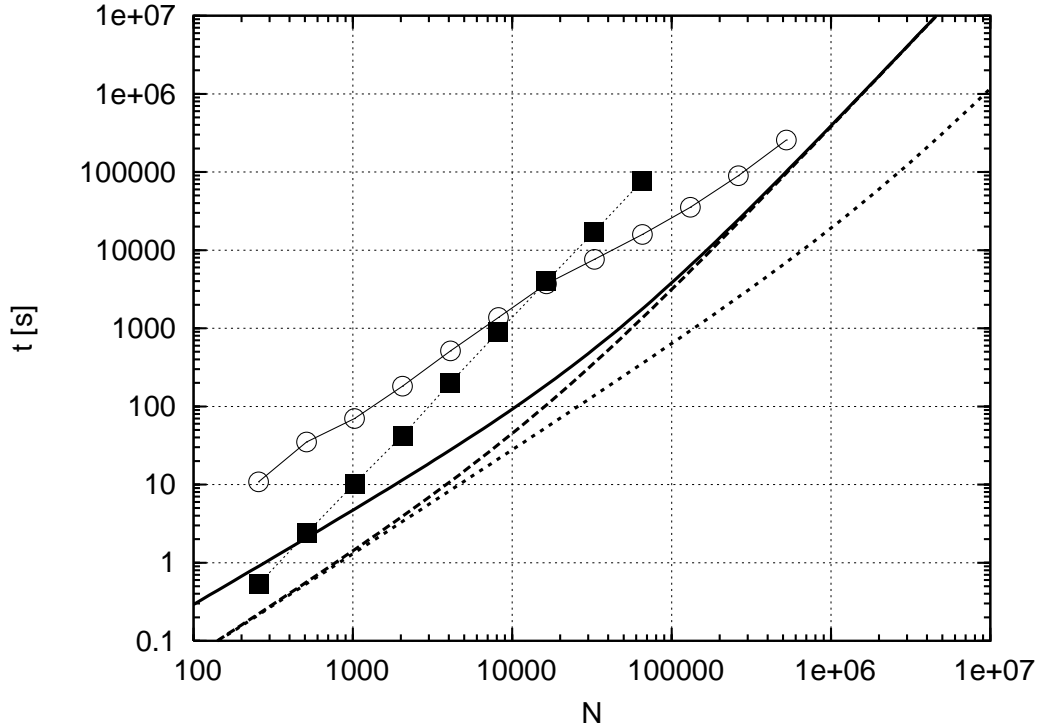,width=\columnwidth}
   \caption[]{Prospective of future CPU and GPU performance, based on
   the model from \S\,\ref{Sect:performance}.  The two thin curves
   with squares and circles give the measured performance of the CPU
   and 8800GTX GPU, respectively. 
   The thick solid curve gives a prediction of the performance for the
   8800GTX in which only blocks of particles are communicated with the
   GPU. The dashed curves gives in addition the effect of carrying the
   predictor/corrector calculation to the GPU.  The doted curve gives
   the performance of a hypothetical 8800GTX-like architecture for
   which in addition the processor pipeline would be used more
   efficiently ($\eta_{\rm fe} = 1$).
\label{fig:Future} }
\end{figure}

These current bottlenecks in the GPU may be reduced using the compute
unified device architecture (CUDA)\footnote{see {\tt
http://developer.nvidia.com/cuda}} programming environment, which is
supposed to provide an improved environment for general purpose
programming on the GPU.  In fig.\,\ref{fig:Future} we present the
possible future performance assuming that the communication additional
overhead on the GPU is lifted, the clock cycles are used more
efficiently without any assumptions of improved hardware speed.  In
the first step we simple reduce communication to blocks rather than
having to transport all particles each block time step (solid
curve). This relatively simple improvement can already be carried out
using CUDA. The second optimalization (dashed curve in
fig.\,\ref{fig:Future}) is achieved when, in addition to reducing the
communication we also carried the predictor and corrector steps to the
GPU. This improvement, however, may be associated with a quite severe
accuracy penalty. For both improvements we used the performance data
for the current design 8800GTX. Further improvement can be achieved
when, in addition to more efficient communication the force pipeline
can be represented more efficiently by the shader pipeline, like is
done on GRAPE. The result of this hypothetical case would improve
performance by more than a factor 100 compared to the workstation over
the entire range of $N$.

\section*{Acknowlegments}

We are grateful to Mark Harris and David Luebke of NVIDIA for
supplying us with the two NVIDIA GeForce 8800GTX graphics cards on
which part of the simulations were performed.  We also like to thanks
Alessia Gualandris and Jun Makino for numerous discussions.  This work
was supported by NWO (via grant \#635.000.303 and
\#643.200.503) and the Netherlands Advanced School for Astrophysics (NOVA).  
The calculations for this work were done on the Hewlett-Packard xw8200
workstation cluster and the MoDeStA computer in Amsterdam, both are
hosted by SARA computing and networking services, Amsterdam.

%\input /home/spz/latex/lib/bib/references
%\bibliographystyle{/home/spz/latex/lib/inputs/aabib}
%\bibliography{gpu}

\section*{Appendix A}

The N-body code presented in this paper consists of a part implemented
in C (running on a CPU) and a part implemented in Cg (running on the
GPU).  In this appendix we show the routine that evaluates the
acceleration, jerk and potential in Cg (which was based on a tutorial
available from \cite{Goeddeke2005b}). The C code which handles
communication between CPU and GPU and supporting data structures is
not presented here.  A copy of the entire working version of the code
is available via {\tt http://modesta.science.uva.nl}.

\newpage
\begin{mylisting}
\begin{verbatim}
void compute_acc_jerk_and_pot(
  in  float2 coords     : TEXCOORD0,      // 2D texture coordinate of this particle
  out float3 acc        : COLOR0,         // Output texture with acceleration 
  out float4 jerkAndPot : COLOR1,         // Output texture with jerk and potential
  uniform samplerRECT accTexture,         // Input texture with all particles' acceleration
  uniform samplerRECT jerkAndPotTexture,  //  ,,     ,,     ,,   ,,    ,,      jerk and potential
  uniform samplerRECT massTexture,        //  ,,     ,,     ,,   ,,    ,,      mass
  uniform samplerRECT posTexture,         //  ,,     ,,     ,,   ,,    ,,      position
  uniform samplerRECT velTexture,         //  ,,     ,,     ,,   ,,    ,,      velocity
  uniform float eps2,                     // Softening parameter
  uniform float otherParticle,            // Index to other particle
  uniform float texSizeX,                 // Width of all textures
  uniform float texSizeY,                 // Height of all textures
  uniform float offset)                   // Number of unused texture elements
{
  float  coords1D, newCoords1D, otherMass,
         r2, xdotv, r2inv, rinv, r3inv, r5inv, xdotvr5inv;
  float2 newCoords;
  float3 pos, otherPos, vel, otherVel, dx, dv, thisAcc, thisJerkAndPot;
  
  // Get data from the textures
  acc        = texRECT(accTexture, coords).rgb;
  jerkAndPot = texRECT(jerkAndPotTexture, coords).rgba;
  pos        = texRECT(posTexture, coords).rgb;
  vel        = texRECT(velTexture, coords).v7.texrgb;

  // Convert the 2D texture coordinate to 1D and increase with otherParticle
  // to obtain the coordinate of this iteration's other particle. Because our
  // textures are defined as samplerRECT, texture elements must be addressed 
  // as (x+0.5,y+0.5). When converting to 1D, we must compensate for this offset).
  coords1D = round(coords.y-0.5)*texSizeX + round(coords.x-0.5);
  newCoords1D = coords1D + otherParticle;

  // Skip over unused texture elements
  if (newCoords1D + offset > texSizeX*texSizeY - 1)
    newCoords1D = newCoords1D - (texSizeX*texSizeY - offset);
  
  // Convert the other particle's 1D coordinate to 2D. As above, we must add
  // 0.5 to obtain correct texture element coordinates.
  newCoords = 0.5 + float2( frac(newCoords1D/texSizeX)*texSizeX,
                            floor(newCoords1D/texSizeX) );

  // Get the position, velocity and mass of this iteration's other particle
  otherPos  = texRECT(posTexture, newCoords).rgb;
  otherVel  = texRECT(velTexture, newCoords).rgb;
  otherMass = texRECT(massTexture, newCoords).r;

  // Compute acceleration, jerk and potential
  dx = otherPos-pos;
  dv = otherVel-vel;
  r2 = eps2 + dot(dx,dx);
  xdotv = dot(dx,dv);
  r2inv = 1.0/r2;
  rinv = sqrt(r2inv);
  r3inv = r2inv*rinv;
  r5inv = r2inv*r3inv;
  xdotvr5inv = 3.0*xdotv*r5inv;
  thisAcc = otherMass*r3inv*dx;
  thisJerkAndPot = otherMass*(r3inv*dv - xdotvr5inv * dx);
  acc = acc + thisAcc;
  jerkAndPot.rgb = jerkAndPot.rgb + thisJerkAndPot;
  jerkAndPot.a = jerkAndPot.a - otherMass*rinv;
}

\end{verbatim}
\end{mylisting}

\end{document}